# The Strong Cell-based Hydrogen Peroxide Generation Triggered by Cold Atmospheric Plasma

Dayun Yan[1*], Haitao Cui[1], Wei Zhu[1], Annie Talbot[2], Lijie Grace Zhang[1], Jonathan H. Sherman[3], and Michael Keidar[1*]

[1]Department of Mechanical and Aerospace Engineering, The George Washington University, Science & Engineering Hall, 800 22nd Street, NW, Washington, DC 20052, USA

[2]Columbian College of Arts and Sciences, The George Washington University, Philips Hall, 801 22nd Street, NW, Suite 212, Washington, DC 20052, USA

[3]Neurological Surgery, The George Washington University, Foggy Bottom South Pavilion, 22nd Street, NW, 7th Floor, Washington, DC 20037, USA

*Corresponding authors: Dayun Yan ydy2012@gwmail.gwu.edu, Michael Keidar keidar@gwu.edu

**Abstract.** Hydrogen peroxide ($H_2O_2$) is an important signaling molecule in cancer cells. However, the significant secretion of $H_2O_2$ by cancer cells have been rarely observed. Cold atmospheric plasma (CAP) is a near room temperature ionized gas composed of neutral particles, charged particles, reactive species, and electrons. Here, we first demonstrated that breast cancer cells and pancreatic adenocarcinoma cells generated micromolar level $H_2O_2$ during just 1 min of direct CAP treatment on these cells. The cell-based $H_2O_2$ generation is affected by the medium volume, the cell confluence, as well as the discharge voltage. The application of cold atmospheric plasma (CAP) in the cancer treatment has been intensively investigated over the past decade. Several cellular responses to the CAP treatment have been observed including the consumption of the CAP-originated reactive species, the rise of intracellular reactive oxygen species, the damage on DNA and mitochondria, as well as the activation of apoptotic events. This is a new previously unknown cellular response to CAP, which provides a new prospective to understand the interaction between CAP and cells.

**Introduction.**

$H_2O_2$ is an important signaling molecule in cancer cells (1). The production of nanomolar (nM) level of $H_2O_2$ by several cancer cell lines including melanomas, neuroblastoma, colon carcinoma, and ovarian carcinoma have been observed two decades ago (2). $H_2O_2$ may increase the genetic instability of cancer cells by inducing DNA strand breaks, damage on guanine or thymine bases, and the sister chromatid exchanges, which may facilitate the malignant process of cancer cells, such as proliferation, apoptosis resistance, metastasis, angiogenesis and hypoxia-inducible factor 1 activation (1, 2). On the other hand, $H_2O_2$ alone with a relative high concentration or as the mediator of a series of anticancer drugs can selectively induce apoptosis in cancer cells (1, 3–5). $H_2O_2$ may have promising application in cancer treatment at least as a mediator of series of physical or chemical strategies.

Cold atmospheric plasma (CAP), a near room temperature ionized gas composed of charged particles, neutral particles and electrons, has shown its promising application in cancer treatment over the past decade (6–11). CAP not only effectively decreases the growth of many cancer cell lines *in vitro* through reactive species-triggered cell death but also significantly inhibits or halts the growth of subcutaneous xenograft tumors or melanoma in mice by the direct CAP treatment just above the skin (8, 14–17). The reactive oxygen species (ROS) and the reactive nitrogen species (RNS) have been regarded as the main factors contributing to the complicate interaction between CAP and cancer cells *in vitro* and *in vivo* (12, 13). Many studies conclude that the death of the CAP-treated cancer cells *in vitro* is due to the apoptosis triggered by the significant rise of intracellular ROS, DNA damage, as well as mitochondrial damage (7, 11, 18–21).

Among dozens of CAP-originated species in aqueous solutions, $H_2O_2$ has been proven to be a main factor triggering the death of cancer cells *in vitro* (22–26). $H_2O_2$ has not been detected in the emission spectra of CAP in the gas phase (19). The $H_2O_2$ in aqueous solution is due to the recombination reactions between short-life species $OH^{.}$ (27, 28). To date, CAP is the only confirmed extracellular $H_2O_2$ source in plasma medicine. Cells have just been regarded as a consumer for the $H_2O_2$ generated by CAP (11, 24). However, using a solution with $H_2O_2$ alone does not cause the same decrease in cancer cells viability as seen following the CAP treatment (29, 30). Thus, the CAP treatment on cancer cells cannot be simply regarded as a $H_2O_2$-treatment.

In addition to the direct CAP treatment, another strategy using the CAP-stimulated solutions (PSS) to inhibit the growth of cancer cells *in vitro* or to inhibit the growth of tumorous tissues in mice through injection has been also demonstrated recently (31–34). PSS is also named as the indirect CAP treatment or the CAP-activated solutions (24, 35). For the direct CAP treatment *in vitro*, a thin layer of cell culture medium is used to cover cancer cells (36). This medium layer facilitates the transition of the reactive species in the gas phase of CAP into the dissolved reactive species in medium (36). The direct CAP treatment can be regarded essentially the same as the indirect CAP treatment if we assume that the finally formed long-life reactive species such as $H_2O_2$, $NO_2^-$, $ONOO^-$ in the CAP-treated solution is the sole factor leading to decreased tumor cell viability. However, considering the potential interaction between the short-life species such as superoxide ($O_2^-$), hydroxyl radicals ($OH^.$) in CAP and cells, the direct CAP treatment should be different from the indirect CAP treatment. PSS will not contain short-life species. So far, no direct experimental evidence has been found to describe such essential difference.

In this study, we first provide the experimental evidence that the 1 min of CAP treatment can trigger µM level cancer cells-based $H_2O_2$ generation through comparing the $H_2O_2$ generation in the CAP-treated cell-free medium and the CAP-treated cancer cells covered by a thin layer of medium. Such cells-based $H_2O_2$ generation can be regulated by controlling the volume of the medium layer, the cell confluence and the discharge voltage. This study provides the first evidence that cells can not only quickly consume $H_2O_2$ but also significantly generate $H_2O_2$ through CAP treatment. This is a novel perspective to understand the interaction between CAP and cells.

**Experiments and Results.**

Consuming the extracellular $H_2O_2$ is a basic response of cancer cells cultured in the $H_2O_2$-containing environment (11, 24). We have demonstrated that the CAP-originated $H_2O_2$ could be completely consumed just several hours after using the CAP-stimulated DMEM to culture glioblastoma (U87MG) cells (11). In this study, we investigated the evolution of extracellular $H_2O_2$ in PSM and $H_2O_2$-DMEM, which have been used to culture cancer cell lines. All solutions contained 36.3 µM $H_2O_2$. The $H_2O_2$ concentration in PSM and $H_2O_2$-DMEM decreases as the culture time increases (Fig. S1). At a higher cell confluence, a faster $H_2O_2$ consumption rate is noted. To date, all three cancer cell lines we studied have shown a similar feature that cancer cells quickly consume the extracellular $H_2O_2$ in just 3 hr. This observation is consistent with the observation that the rise of intracellular ROS started immediately after the CAP treatment and ended just about 3 hr post the CAP treatment (37).

To date, the direct CAP treatment is the main strategy to investigate the toxicity of CAP on cancer cells *in vitro* (8, 31, 32). In this study, we measured the $H_2O_2$ concentration in a thin layer of DMEM which has been used to immerse cells during the direct CAP treatment. Without the protection of such medium layer, cancer cells are killed immediately due to the dehydration caused by the helium flow from the CAP jet tube. Based on the preliminary test, 20 µL per well on a 96-wells plate was the minimum volume to preclude the impact of helium flow (Fig. S2). In addition, the volume effect of this thin DMEM layer was also investigated through increasing the volume from 20 μL to 100 μL. Three general trends have been observed when we immediately measured the $H_2O_2$ concentration in DMEM after the treatment. First, the concentration of $H_2O_2$ in all cases increases as the volume of DMEM decreases (Fig. 1A). This result may be due to the principle that a salute will form in a larger concentration when the volume of solvent (reactive species) is smaller. Second, the direct CAP treatment on pancreatic adenocarcinoma cells (PA-TU-8988T) and breast adenocarcinoma cells (MDA-MB-231) rather than on glioblastoma cells (U87MG) generates significantly more $H_2O_2$ than the same CAP treatment on DMEM in the same volume (Fig. 1A). This trend is more obvious when the volume of DMEM is smaller. U87MG cells will show similar feature only at some specific conditions such as at a medium volume of 80 μL and 100 μL. The generation of $H_2O_2$ is also illustrated as the change in $H_2O_2$ concentration between the CAP-treated DMEM with and without cells (Fig. 1B). The volume of DMEM is also a key factor to affect the cells-based $H_2O_2$ generation. The maximum relative $H_2O_2$ generation from MDA-MB-231 cells, PA-TU-8988 cells, and U87MG occurs when the volume of DMEM is 50 μL, 100 μL, and 80 μL, respectively (Fig. 1B). More importantly, only the direct CAP treatment on cancer cells can cause such cells-based $H_2O_2$ generation. The CAP-stimulated DMEM cannot generate the similar response on all three cell lines (Fig. 1C). The response of cancer cells to PSM is just consuming $H_2O_2$, as seen in Fig. S1. In addition, all three cell lines do not generate $H_2O_2$ without the CAP treatment.

To investigate whether the CAP-treated cancer cells will continue to generate $H_2O_2$ post the CAP treatment, the change of $H_2O_2$ concentration in DMEM after the direct CAP treatment on cancer cells and the CAP-treated DMEM has been studied. For all cell lines, the $H_2O_2$ concentration in DMEM gradually decreases post the CAP treatment (Fig. 2). Such a decay trend has not been observed during the evolution of the CAP-treated DMEM. Thus, the decreased $H_2O_2$ concentration is due to the consumption of cancer cells on the extracellular $H_2O_2$. The cells-based $H_2O_2$ generation occurs only when the direct CAP treatment is performed. Once the CAP treatment ceases, the generation of $H_2O_2$ stops. When the $H_2O_2$ concentration measurement was performed at the tenth min after the CAP treatment, the cells-based $H_2O_2$ generation was no longer observed (Fig. 2). This may explain why the cells-based $H_2O_2$ generation has not been observed in the previous studies.

In addition to the volume of DMEM used to cover cancer cells during the CAP treatment, the cell confluence and the discharge voltage can also affect the cells-based $H_2O_2$ generation. Under the same experimental conditions, the maximum of the cells-based $H_2O_2$ generation of MDA-MB-231 cells, and PA-TU-8988T cells appears when the cell confluence is $1 \times 10^4$ cells/mL and $3 \times 10^4$ cells/mL, respectively (Fig. 3A). For U87MG cells, the significant effect of the cell confluence on the cells-based $H_2O_2$ generation was not observed (Fig. 3A). The discharge voltage is an important physical factor during the CAP treatment. The change of discharge voltage in a short range can significantly change the chemical composition of CAP. According to prior research, the increase of output (discharge) AC voltage from 2.56 kV to 3.80 kV will significantly increase the generation of the nitrogen-based species such as $N_2^+$ and $NO/N_2$ in the gas phase of CAP (38). The generation of the oxygen-based species such as O and $OH^.$ just slightly increases during the same process (38). Thus, the ratio between the oxygen-based species and the nitrogen-based species decreases as the output voltage increases. We observed the same trend in the CAP-stimulated PBS that the concentration ratio between $H_2O_2$ and $NO_2^-$ decreases from 11.2 to 5.5 as the output voltage increases from 3.02 kV to 3.85 kV. In this study, the increase of the output voltage significantly weakens the cells-based $H_2O_2$ generation (Fig. 3B). A large output voltage will completely inhibit the cells-based $H_2O_2$ generation for all three cell lines.

$H_2O_2$ has been regarded as a main anti-cancer reactive species in the CAP treatment *in vitro* (24–26). Because the observation that the direct CAP treatment on cancer cells immersed in a thin layer of DMEM causes a significantly stronger $H_2O_2$ generation than that seen with the indirect CAP treatment on DMEM, it is reasonable to speculate that the direct CAP treatment will cause a stronger anti-cancer effect on cancer cells than the indirect CAP treatment does under the same experimental conditions. Such a comparison was performed on MDA-MB-231 cells and PA-TU-8988T cells. As shown in Fig. 4A and 4B, both cell lines are much more vulnerable to the direct CAP treatment than the indirect CAP treatment. In addition, it was further demonstrated that the strong anti-cancer effect of the direct CAP treatment on the two cell lines could nearly be completely counteracted by pre-treating cancer cells with 6 mM intracellular ROS scavenger NAC or by removing the DMEM immersing cancer cells and replacing it with new untreated DMEM immediately after the direct CAP treatment (Fig. 5C and 5D). Thus, the anti-cancer effect of the direct CAP treatment on cancer cells is still based on a rise of intracellular ROS due to the extracellular reactive species dissolved in DMEM. Moreover, the volume of DMEM does not show a noticeable impact on the anti-cancer effect of the direct or the indirect CAP treatment, though corresponding $H_2O_2$ concentration in DMEM varies drastically with the volume (Fig. 1). This difference indicates that other factors such as the whole reactive species amount rather than the concentration of reactive species may dominate the anti-cancer effect. In addition to the general trend introduced above, a slight difference on the anti-cancer effect due to the volume of DMEM still can be observed on MDA-MB-231 cells. The strongest anti-cancer effect of the direct CAP treatment on MDA-MB-231 cells occurs when the volume of DMEM is 50 μL.

**Discussion.**

Since the first observation that the CAP treatment could cause apoptosis in lung cancer cells line in 2004 and the first complete study on the anti-cancer effect of dielectric barrier discharge (DBD) on melanoma cells in 2007, the research on the application of CAP in the cancer treatment experienced an exponential development (7, 8, 11). While understanding the anti-cancer mechanism at the molecular and cellular level is far from clear, it is widely acknowledged that the CAP-originated reactive species are the main anti-cancer factors inhibiting the growth of cancer cells *in vitro* (6–8, 10, 11). However, in all previous studies, the cancer cells have just been regarded as having a passive role in plasma medicine to sustain the attack from the reactive species such as $H_2O_2$ (6–8, 10, 11). This study first demonstrates that specific cancer cells such as pancreatic adenocarcinoma cells and breast adenocarcinoma cells will generate significant amount of $H_2O_2$ as a response to the direct CAP treatment. For example, MDA-MB-231 cells generate about 85% more $H_2O_2$ than the CAP treatment does when the volume of DMEM was 50 μL. Despite only three cell lines has shown such cells-based $H_2O_2$ generation capacity in this study, such cellular response may be commonly existed in other cancer cell lines which has not been studied. And, it is completely possible that such feature may also be owned by other cells including normal mammalian cells, bacteria, as well as yeasts. These cells have been investigated in plasma medicine for decades (37, 39, 40).

We hypothesize that the short-life species in CAP may stimulate the $H_2O_2$ production by cancer cells. First, the $H_2O_2$ generation by cancer cells occurs only during the CAP treatment. The cessation of CAP treatment will immediately cause the consumption of $H_2O_2$ to be the dominate response of cancer cells (Fig. 2). Consuming $H_2O_2$ is a basic response of cancer cells in an $H_2O_2$-containing environment (Fig. S1). It is possible that $H_2O_2$ consumption capacity and the $H_2O_2$ generation capacity coexist during the CAP treatment. U87MG cells show a much stronger $H_2O_2$ consumption capacity compared with breast cancer cell lines (11, 24). Thus, the weak cell-based $H_2O_2$ generation by U87MG cells may be due to its strong $H_2O_2$ consumption capacity which dominates the interaction between U87MG cells and CAP.

The short-life species in CAP has been regarded as a main source to form $H_2O_2$ in the CAP-stimulated solutions, though the corresponding mechanism is still disputable (41, 42). Such $H_2O_2$ formation is an aqueous solution-based reaction. Hydroxyl radicals may form $H_2O_2$ by a simple reaction: $OH^{.} + OH^{.} \rightarrow H_2O_2$ (27, 28). $H_2O_2$ may also be formed by the recombination reaction based on hydroperoxyl radicals ($HO_2$): $HO_2 + HO_2 \rightarrow H_2O_2 + O_2$ (27). $HO_2$ may be formed by a reaction involving $O_2^{-}$: $O_2^{-} + H^{+} \rightarrow HO_2$ (27). For the cells-based $H_2O_2$ generation, $H_2O_2$ may be formed by a dismutation reaction catalyzed by the extracellular superoxide dismutase (Ex-SOD, SOD3) on the cytoplasmic membrane of cancer cells (22, 23). Superoxide is provided by CAP (39). The expression of SOD3 have been found in human pancreatic adenocarcinoma tissues, estrogen-induced breast cancer tissues, and breast carcinoma cell (MCF-7, MDA-MB-231), but not in glioblastoma cells, which may explain why cell-based $H_2O_2$ production has not been noticeably observed in the CAP-treated U87MG cells (43–45).

The discovery of cell-based $H_2O_2$ generation provides a new feature which distinguishes the CAP treatment and simple chemical treatment such as the $H_2O_2$ treatment. The direct CAP treatment causes a much stronger anti-cancer effect on pancreatic adenocarcinoma cells and breast adenocarcinoma cells than the indirect CAP treatment does (Fig. 4). Such difference may be at least partially due to the cells-based $H_2O_2$ generation during the CAP treatment (Fig. 1). Nonetheless, it is necessary to emphasize that the weak anti-cancer capacity of the indirect CAP treatment (PSS) revealed in this study is mainly due to the small diameter of the wells in 96-wells plate. As we demonstrated in previous study, the diameter of multi-wells plate is proportional to the reactive species concentration and the anti-cancer effect of PSS (24). Consequently, the PSS made in 6-wells plate will be much more toxic to cancer cells than the PSS made in 96-wells plate (24).

More importantly, this study provides a new perspective to understand the potential cellular interaction during the CAP treatment, which also provide clues to understand the anti-cancer capacity of CAP *in vivo*. In addition to the proven selective anti-cancer capacity of CAP *in vitro* to cancer cells, another attractive feature of CAP is its promising anti-cancer effect seen *in vivo*. A series of investigations have achieved a similar conclusion that the growth of subcutaneous xenograft tumors or melanoma in mice can be significantly halted by the CAP treatment just above the skins of mice or on the exposed cancerous tissues in mice (15–17, 46). The mechanism governing these descriptions is nearly completely unknown. ROS and RNS may also be the main factors contributing to the decreased tumor cell viability *in vivo* by the CAP treatment through directly attacking tumor or indirectly activating the immune response *in vivo* to further kill tumor cells (18, 47, 48). The trans-skin motion (diffusion, transportation or other physical ways) of reactive species may be a key to understand the anti-cancer capacity *in vivo*. Several attempts have been performed to understand such processes. The diffusion of reactive species across the skin analogue has been observed (49, 50). Our study provides a novel explanation for the trans-skin motion of reactive species. The short-life reactive species, may activate cells to generate long-life reactive species such as $H_2O_2$ through the similar ways revealed in this study. The tumor tissue may be immersed in a $H_2O_2$-rich environment even the CAP treatment does not directly generate a significant amount of $H_2O_2$ on the relative dry skin. The cells in the skin and the cells in the tumor may have similar mutual interaction through generating toxic chemicals such as $H_2O_2$ to neighbor cells. $H_2O_2$ is a second messenger in the lymphocyte activation (51, 52). µM level of $H_2O_2$ rapidly induced the activation of the transcription factor NF-κB and early gene expression of interleukin-2 (IL-2) and the IL-2 receptor α chain (51). The $H_2O_2$-generating tumor tissues may be a potential target for the immune system. The anti-cancer capacity of the CAP treatment *in vivo* may involve the $H_2O_2$-activated immune attack on tumor tissues.

**Conclusions.**

A new previously unknown basic cellular response to the CAP treatment is demonstrated in this study. Only direct CAP treatment on breast adenocarcinoma cells and pancreatic adenocarcinoma cells immersed in a thin layer of medium results in µM level of cell-based $H_2O_2$ generation. The measured maximum $H_2O_2$ generation based on CAP-stimulated MDA-MB-231 cells immersed in a thin layer of DMEM is about 85% more than that formed in the CAP-stimulated same medium but lacking cells. Controlling the volume of medium, the cell confluence, and the plasma discharge voltage can regulate the cell-based $H_2O_2$ generation. The abundant short-life reactive species in CAP may trigger this unique cellular response, which gives a new perspective to understand the interaction between CAP and cells *in vitro* and *in vivo*.

**Methods.**

**CAP device**. The CAP device used in this study was a typical CAP jet generator using helium as the carrying gas. The noticeable anti-cancer effect of this device has been demonstrated through a series of previous investigations from our lab (24, 53). The detailed introduction for this device was illustrated in previous reports (24, 53). Here, a short introduction is given. A violet plasma jet was generated between a central anode and a ring grounded cathode. The discharge was driven by an alternating current high voltage (3.16 kV) with a frequency of 30 kHz. The generated CAP was ejected out from a quartz tube with a diameter of 4.5 mm. The flow rate of helium was about 4.7 L/min. The input voltage of DC power was 11.5 V. According to the emission spectrum, the CAP in the gas phase was mainly composed of ROS (OH·, O), RNS (NO, $N_2^+$), and helium (He) (38).

**Cell cultures**. Human pancreas adenocarcinoma cells (PA-TU-8988T) and human glioblastoma cells (U87MG) were provided by Dr. Murad's lab at the George Washington University. Human breast adenocarcinoma cells (MDA-MB-231) were provided by Dr. Zhang's lab at the George Washington University, and cultured in the same protocol as previous studies. The Dulbecco's modified Eagle's medium (DMEM) was purchased from Life Technologies (11965-118). DMEM was mixed with 1% (v/v) penicillin and streptomycin solution (Life Technologies, 15140122). The medium used in the cell culture and seeding was composed of DMEM supplemented with 10% (v/v) fetal bovine serum (ThermoFisher Scientific, 26140079) and 1% (v/v) penicillin and streptomycin solution (Life Technologies, 15140122). In each experiment, 100 μL of cancer cells harvesting solutions were seeded in a well on a 96-wells plate (Corning, 62406-081) a day prior to the CAP treatment. These cancer cells were cultured 24 hr under the standard cell culture conditions (a humidified, 37°C, 5% $CO_2$ environment). All wells on the margins of 96-wells plate were not used.

**Making (N-Acetyl-L-cysteine) NAC-DMEM and pre-treating cancer cells.** 6 mM NAC-DMEM was made by dissolving NAC powder (A7250, Sigma-Aldrich) in DMEM. The whole medium which has been used to culture cancer cells overnight were removed first. Then, 100 μL of NAC-rich DMEM was used to culture cancer cells in the well of a 96-wells plate. After 3 hr, 6 mM NAC-DMEM was removed before the further treatment.

**The direct and the indirect CAP treatment on cancer cells.** The gap between the bottom of 12-wells plate and the CAP source was set to be 3 cm. All DMEM used in the CAP treatment was made of mixing 1% (v/v) penicillin-streptomycin solution (Life Technologies, 15140122) with standard DMEM (Life Technologies, 11965-118). The schematically illustration of the protocols was presented in Fig. S4. Prior to the direct CAP treatment, the medium which has been used to culture cells overnight was removed. After this step, for the direct CAP treatment, the DMEM with a specific volume such as 20 μL was transferred to cover the cancer cells in a well on a 96-wells plate. After that, the CAP jet was used to vertically treat the cancer cells for 1 min. For the indirect CAP treatment, the CAP-stimulated DMEM (PSM) was used to affect the growth of cancer cells cultured in a 96-wells plate. To make PSM, the DMEM with a specific volume such as 30 μL in a well on a 96-wells plate was treated by CAP jet vertically for 1 min.

Such PSM was then transferred to affect the cancer cells which have been cultured overnight. For both direct and indirect CAP treatment, the control group corresponded to the cancer cells grown in the new DMEM with a specific volume without the CAP treatment. Then, cancer cells were cultured in the incubator under the standard culture conditions for 3 hr. Subsequently, the thin layer of DMEM around the cancer cells in each well was removed. 100 μL of new untreated DMEM was added in the well and cultured cancer cells for 3 days before the final cell viability assay. In this study, the treatment just using helium rather than CAP was also performed. The protocols were the same as the description above except that no discharge was used. In another case, the DMEM covering the cells needed to be replaced immediately after the CAP treatment. After removing the CAP-treated DMEM, 100 μL of new untreated DMEM was added to culture cells for 3 days before the final cell viability assay.

**Cell viability assay.** MTT (3-(4,5-Dimethyl-2-thiazol)-2,5-Diphenyl-2H-tetrazolium Bromide) assay was performed following the protocol provided by manufacturer (M2128, Sigma-Aldrich). The original experimental data about the cell viability was the absorbance at 570 nm measured by a H1 microplate reader (Hybrid Technology). The original absorbance at 570 nm was processed to be a relative cell viability through the division between the absorbance of the CAP-treated cancer cells and the absorbance of the cancer cells without the CAP treatment.

**Extracellular $H_2O_2$ assay**. The $H_2O_2$ concentration in DMEM or PBS was measured by using Fluorimetric Hydrogen Peroxide Assay Kit (Sigma-Aldrich, MAK165-1KT) following the protocols provided by manufacturer (Sigma-Aldrich). The fluorescence was measured by a H1 microplate reader (Hybrid Technology) at 540/590 nm. The final fluorescence of the experimental group was obtained by deducting the measured fluorescence of the control group from the measured fluorescence of the experimental group. Based on the standard curve, the $H_2O_2$ concentration in DMEM or PBS was obtained. Because the recommended volume of the sample solution was 50 μL, when the volume of sample solution was less than 50 μL but larger than 10 μL, we just collected 10 μL of the sample solution and mixed it with 40 μL of the untreated DMEM to make a 50 μL of the diluted sample solution. The $H_2O_2$ concentration of such case was 5 times larger than the measured concentration.

**Measuring the $H_2O_2$ consumption speed by cancer cells.** The protocols for different cancer cell lines are the same. Here, we just use PA-TU-8988T cells as an example. First, 100 μL of the PA-TU-8988T cells harvesting solution was seeded in a well on a 96-wells plate with a confluence of $6 \times 10^4$ cells/mL. 3 wells were seeded as 3 samples for one experiment. Cells were cultured in incubator for 6 hr under the standard conditions. Then, 1 mL of DMEM or PBS in a well on a 12-wells plate was treated by CAP for 1 min. After that, the medium which has been used to culture cells for 6 hr was removed. 120 μL of the CAP-treated DMEM or $H_2O_2$-containing DMEM ($H_2O_2$-DMEM) was transferred to culture PA-TU-8988T cells in one well on a 96-wells plate. 36.3 μM $H_2O_2$-DMEM was made by adding 9.8 M $H_2O_2$ standard solution (216763, Sigma-Aldrich) in DMEM. Since then until the third hour, 50 μL of the DMEM which has been used to culture cells was transferred to a well on a black clear bottom 96-wells plate in triplicate every hour. Ultimately, the residual $H_2O_2$ concentration in DMEM was measured using fluorimetric hydrogen peroxide assay kit illustrated above.

**Acknowledgements.** This work was supported by National Science Foundation, grant 1465061.

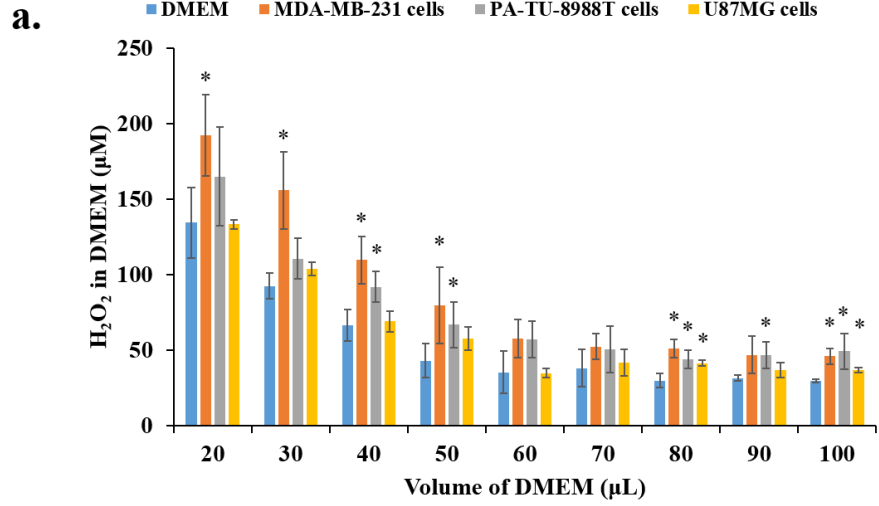

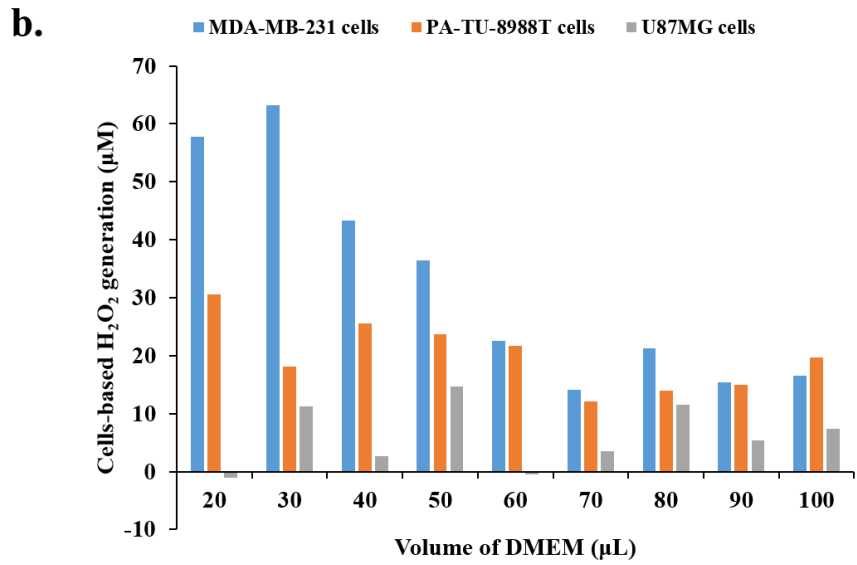

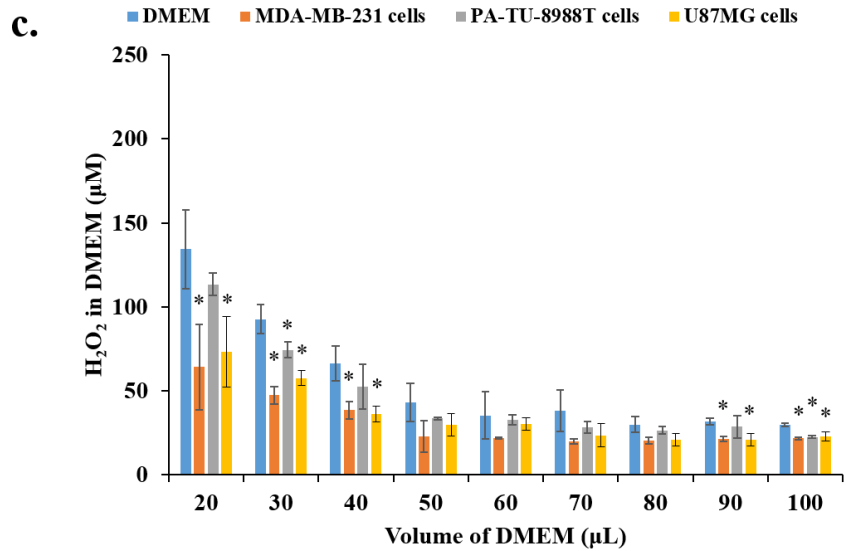

**Fig. 1. Only the direct CAP treatment can trigger the $H_2O_2$ production of specific cancer cell lines *in vitro*.** (A) The $H_2O_2$ concentration in DMEM after the CAP treatment on just DMEM, on pancreatic adenocarcinoma cells (PA-TU-8988T) immerged in DMEM, on breast adenocarcinoma cells (MDA-MB-231) immerged in DMEM, as well as on glioblastoma cells (U87MG) immerged in DMEM. (B) The cells-based $H_2O_2$ concentration. The data is calculated based on the following formula. Cells-based $H_2O_2$ concentration = $H_2O_2$ concentration in the DMEM which has been used to immerse cancer cells during the CAP treatment – $H_2O_2$ concentration in the CAP-treated DMEM. The $H_2O_2$ concentration corresponds to the mean of each bar shown in 1A. (C) The indirect CAP treatment will not stimulate cancer cells to generate $H_2O_2$. All $H_2O_2$ concentration was measured immediately (about 1 min) after the treatment. Results are presented as the mean ± s.d. of three independently repeated experiments. Student's t-test was performed between the data based on cells immersed in specific volume of DMEM and the data just based on the DMEM with the same volume. The significance is indicated as * $p < 0.05$.

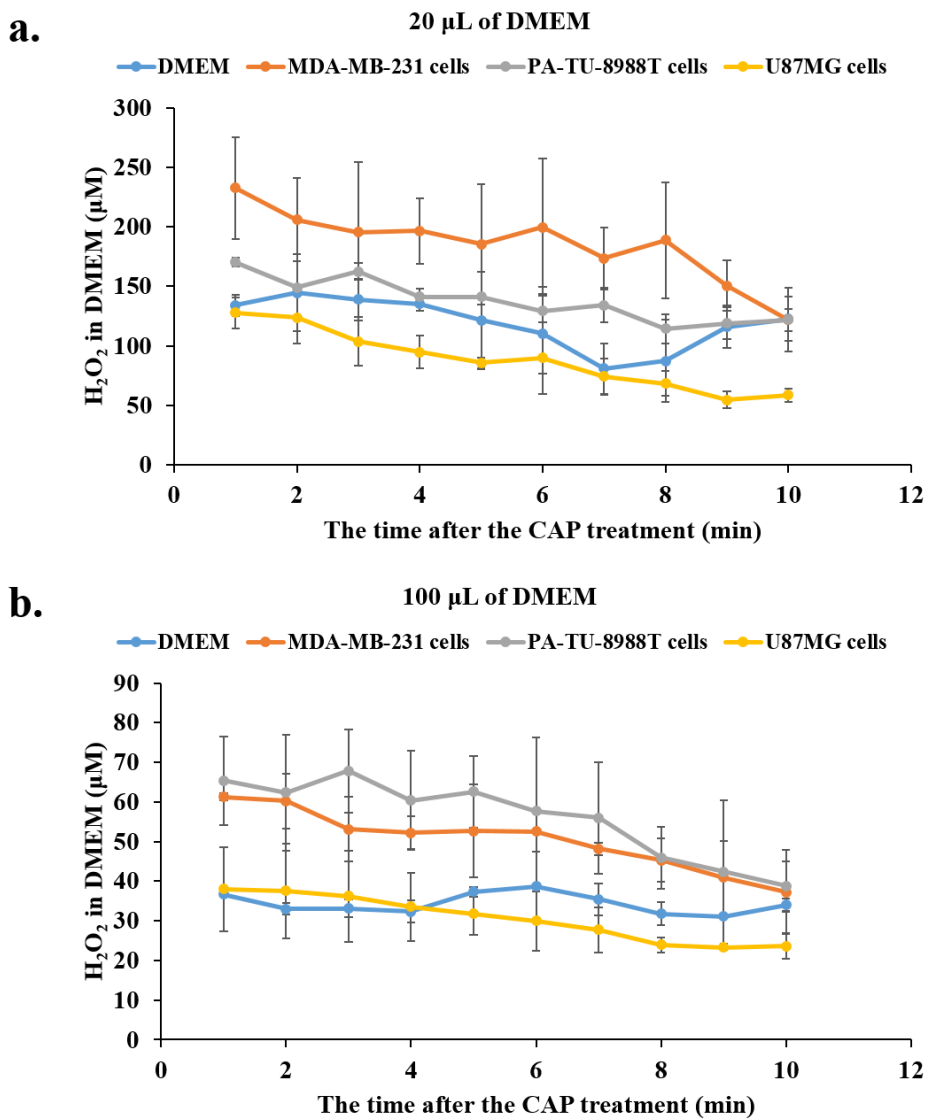

**Fig. 2. H₂O₂ evolution in the DMEM surrounding cancer cells after the direct CAP treatment.** (a) Cancer cells immersed in 20 μL of DMEM. (b) Cancer cells immersed in 100 μL of DMEM. CAP was used to treat DMEM or cancer cells (MDA-MB-231 cells, PA-TU-8988T cells, and U87MG cells) immersed in DMEM. In each experiment, the 1 min of CAP treatment was performed on the cells immersed in 20 μL or 100 μL of DMEM. The H₂O₂ concentration was measured very minute after the CAP treatment until the tenth minute. Cancer cells in 96-wells plate were cultured overnight before the CAP treatment. The cell confluence was 3 x

10⁴ cells/mL. Results are presented as the mean ± s.d. of three independently repeated experiments.

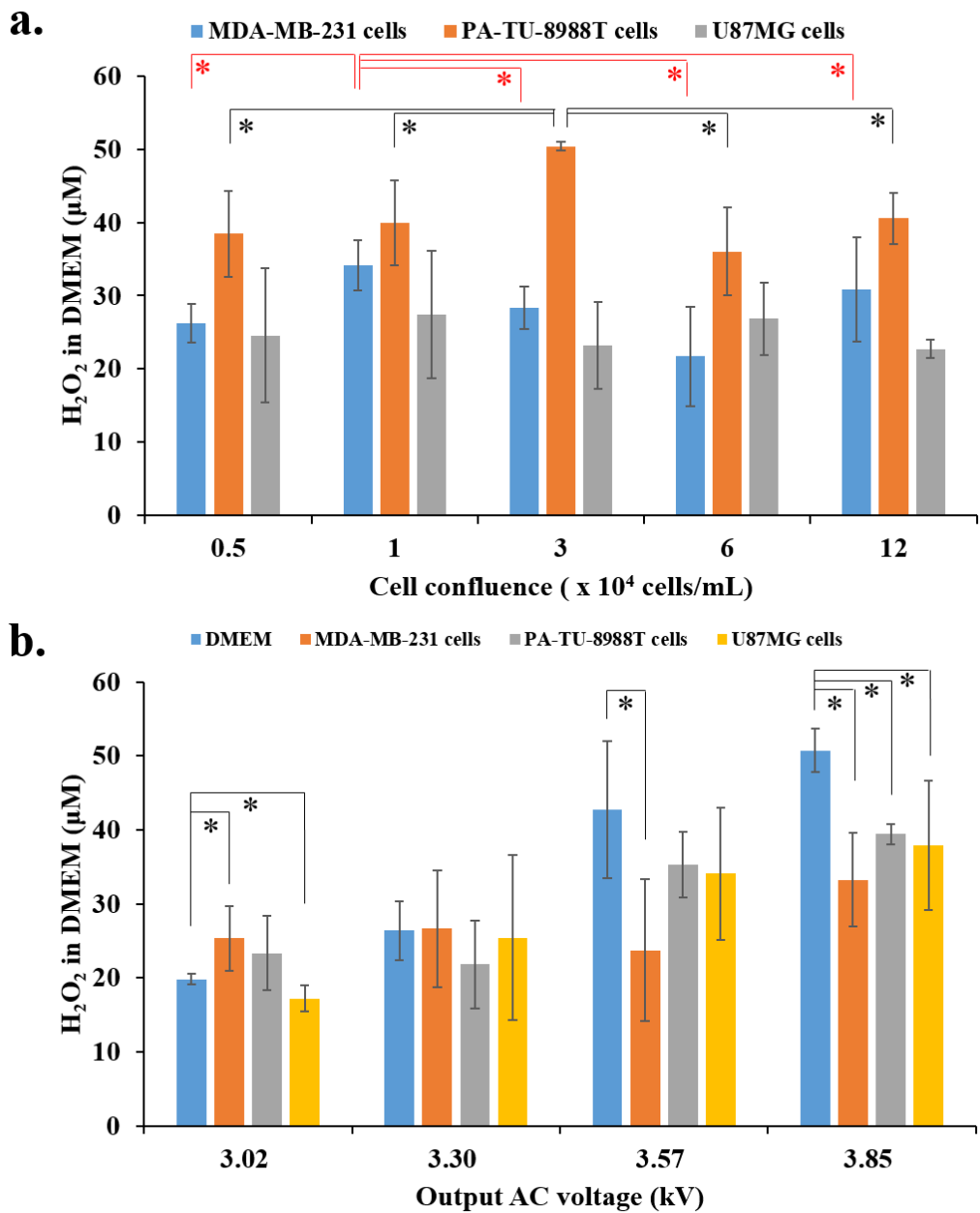

**Fig. 3. The cell confluence and the output voltage affect the cell-based H$_2$O$_2$ generation.** (a) A relative small confluence can achieve the maximum cell-based H$_2$O$_2$ generation. (b) High output AV voltage significantly weakens the cell-based H$_2$O$_2$ generation. In each experiment, a 1

min of CAP treatment was performed on the cancer cells immersed in 100 μL of DMEM. The $H_2O_2$ concentration was measured immediately (about 1 min) after the treatment. Cancer cells were cultured overnight before the treatment. The cell confluence for (b) was $3 \times 10^4$ cells/mL. Results are presented as the mean ± s.d. of three independently repeated experiments. Student's t-test was performed and the significance is indicated as * $p < 0.05$.

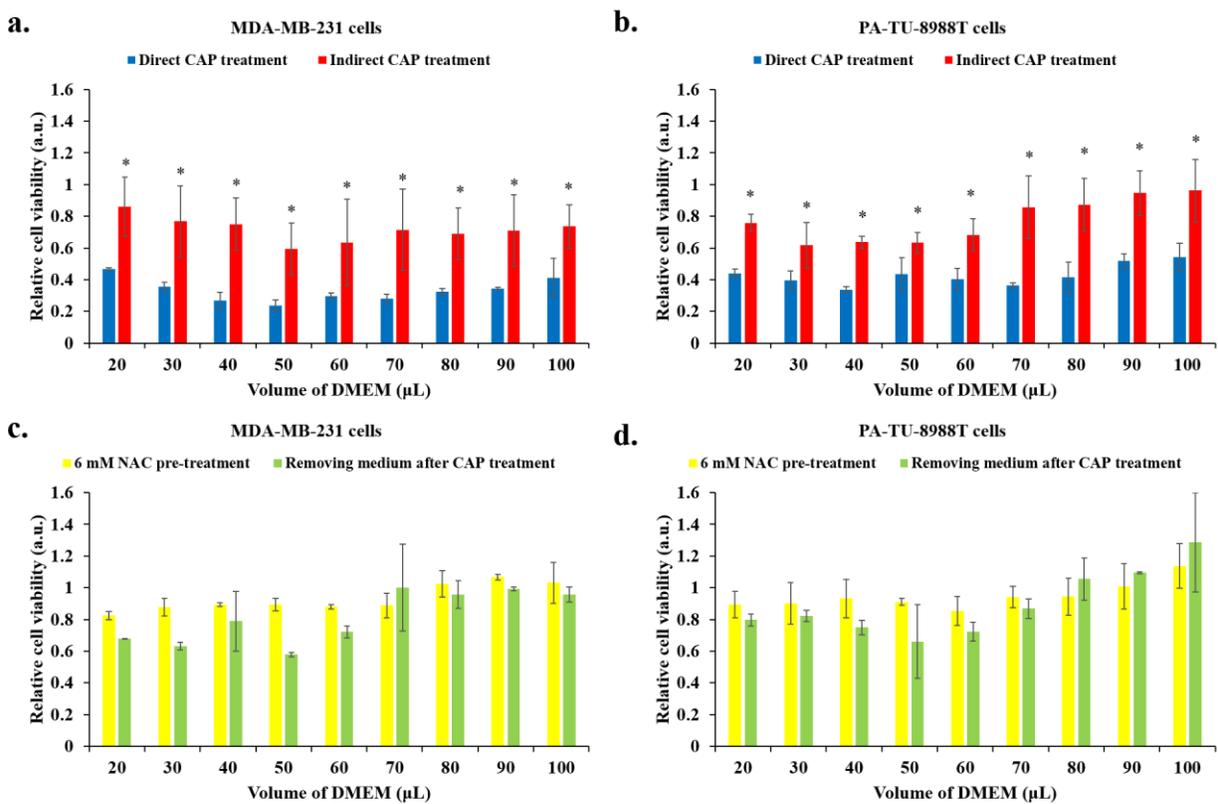

**Fig. 4. Under the same experimental conditions, the anti-cancer capacity of the direct CAP treatment is significantly stronger than the indirect CAP treatment.** For the indirect CAP treatment, the CAP-stimulated DMEM was used to affect the growth of cancer cells. Comparison between the direct CAP treatment and the indirect treatment is performed on (a) MDA-MB-231 cells and (b) PA-TU-8988T cells. Pre-treating cancer cells with 6 mM NAC an intracellular scavenger for 3 hr or immediately (about 1 min) renewing the DMEM after the direct CAP

treatment can effectively weakens the anti-cancer capacity of the direct CAP treatment on (c) MDA-MB-231 cells and (d) PA-TU-8988T cells. Results are presented as the mean ± s.d. of three independently repeated experiments. Student's t-test was performed between the results from the direct CAP treatment and the results from the indirect CAP treatment. The significance is indicated as * $p < 0.05$.

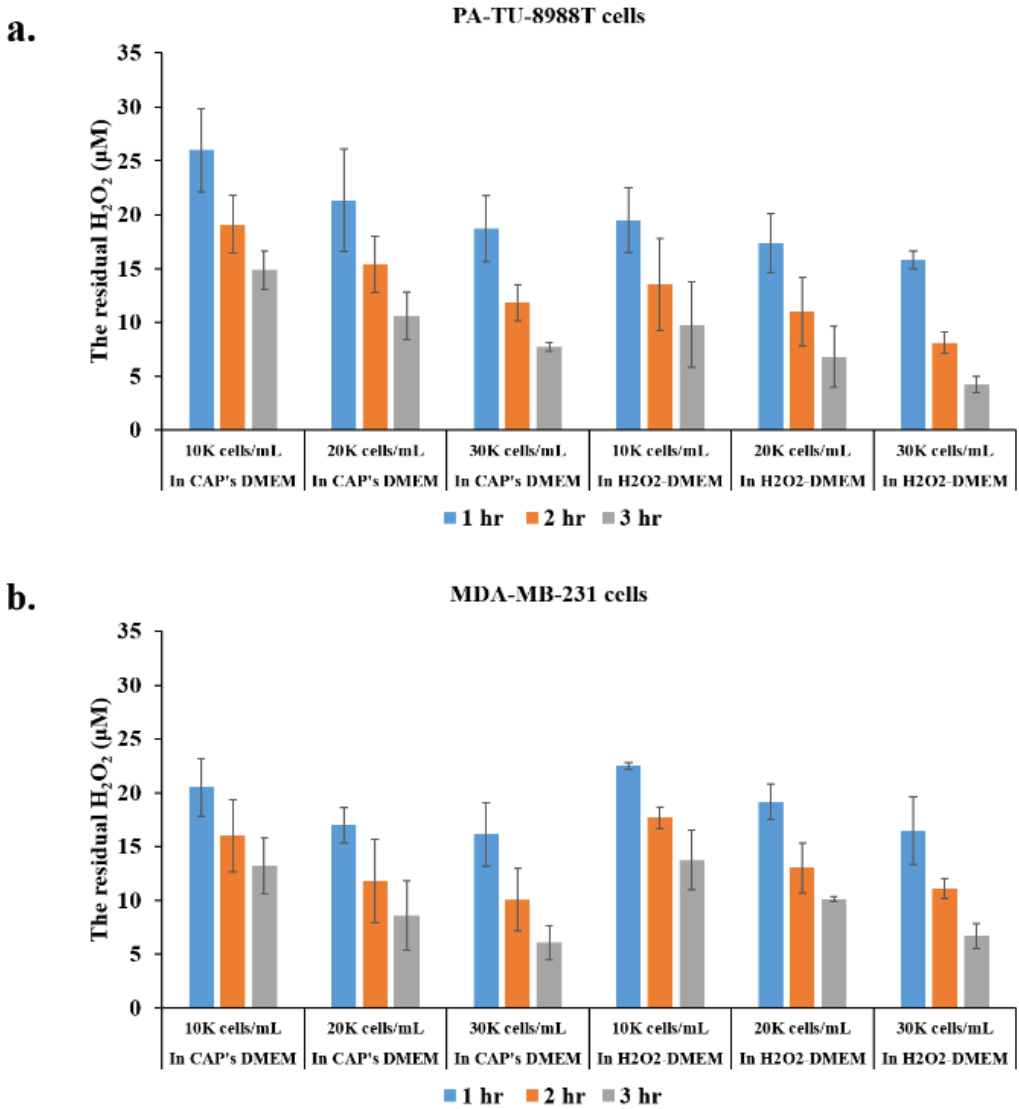

**Fig. S1. Cancer cells quickly consume the $H_2O_2$ in CAP-stimulated DMEM (CAP's medium) and in $H_2O_2$-containing DMEM ($H_2O_2$-DMEM) in hours.** Consuming $H_2O_2$ is a basic feature of cancer cells exposed to $H_2O_2$. 1 hr, 2 hr, and 3 hr represent the time length that two cell lines cultured in CAP's DMEM or in $H_2O_2$-DMEM. The initial $H_2O_2$ concentration in CAP's medium and $H_2O_2$- DMEM is the same and is not shown at here. K represent $1 \times 10^3$. Results are presented as the mean ± s.d. of two independently repeated experiments in triplicate.

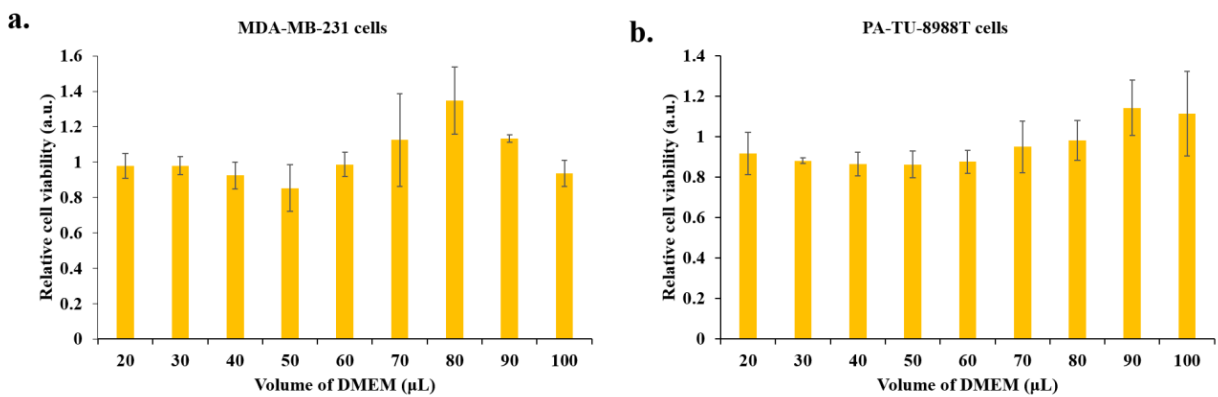

**Fig. S2. The effect of helium flow on the cell viability of cancer cells.** (a) PA-TU-8988T cells. (b) MDA-MB-231 cells. Results are presented as the mean ± s.d. of three independently repeated experiments.

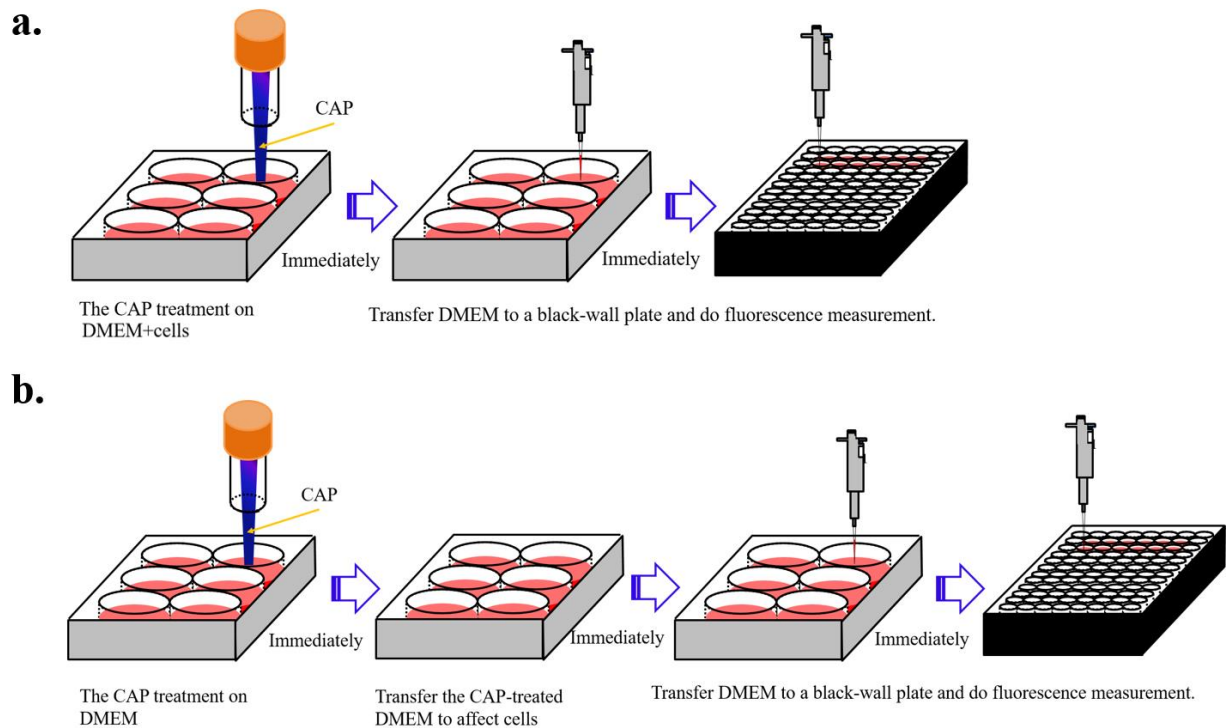

**Fig. S3. Schematic illustration of the protocols for the CAP treatment.** (a) Direct CAP treatment. (b) Indirect CAP treatment.